\documentclass[pra,aps,amssymb,amsmath,amsmath,showpacs,reprint,twocolumn]{revtex4-1}
\usepackage{color}
\usepackage{graphicx}
\usepackage{epstopdf}
\usepackage{natbib}
\usepackage{hyperref}
\usepackage{dcolumn}
\usepackage{young}
\usepackage[aligntableaux=top]{ytableau}
\usepackage[]{youngtab}
\usepackage{bm}
\usepackage{dcolumn}
\newcolumntype{d}[1]{D{.}{.}{#1}}

\newcommand{\br}{\bm{r}}

\newcommand{\ba}{\bm{a}}
\newcommand{\bb}{\bm{b}}
\hypersetup{%
   pdfpagemode=None, 
   pdfstartpage=1,
   pdfstartview=FitH,
   pdfmenubar=true,
   pdftoolbar=true,
   colorlinks = true,
   linkcolor=blue,
   citecolor=blue,
   bookmarksopen=false
}

\newcommand{\Fkt}[1]{\,\mathsf {#1}}
\def\openone{\leavevmode\hbox{\small1\kern-3.3pt\normalsize1}}

\ifx\Tr\renewcommand{\Tr}{\Fkt{Tr}} 
\else\newcommand{\Tr}{\Fkt{Tr}}
\fi

\usepackage{booktabs}
\usepackage{graphicx}
\usepackage{amsmath}
\usepackage{indentfirst}
\begin{document}
\title{Size consistency and counterpoise correction in explicitly correlated calculations\\ of 
interaction energies and interaction-induced properties}

\author{\sc Micha\l\ Lesiuk}
\email{e-mail: lesiuk@tiger.chem.uw.edu.pl}
\author{\sc Bogumi\l\ Jeziorski}
\affiliation{\sl Faculty of Chemistry, University of Warsaw\\
Pasteura 1, 02-093 Warsaw, Poland}
\date{\today}

\begin{abstract}
Explicitly correlated calculations of interaction energies with wave  functions that include all interparticle
distances have suffered so far from the
lack of size-consistency resulting from the difficulty to define monomer
energies corresponding to the applied dimer basis. As a consequence it has not
been possible to obtain interaction energies vanishing at infinite
intermonomer distance  $R$. This has dramatically reduced the accuracy of
calculations at distances where the error in the dimer energy was
comparable with the interaction energy itself. The same problem occurs in
calculations of interaction-induced properties. In this communication we show
how to circumvent this difficulty and obtain interaction energies or
interaction-induced properties that vanish at large $R$. This is
achieved by relaxing the Pauli principle in the diagonalization  of the
 Hamiltonian of noninteracting monomers. The basis functions used for this
diagonalization belong to the representation of the permutation group of the 
dimer
induced by the product of representations appropriate for the monomer spin
states. Nonlinear parameters of the basis set are optimized only for the dimer
in the Pauli-allowed sector of the Hilbert space. In this way, one obtains
$R$-dependent energy of noninteracting monomers  and the corresponding 
interaction
energy includes a counterpoise correction for the basis set superposition
error. The efficiency of this procedure is demonstrated for the
interaction of two hydrogen atoms where accurate reference data are known. 

\end{abstract}

\maketitle

\ytableausetup{boxsize=1.0em}
\ytableausetup{aligntableaux=top}
\ytableausetup{nobaseline}

\section{Introduction}
\label{sec:intro}

In many applications of   electronic structure theory one is concerned with 
changes of a certain 
property of the system resulting from interactions with other atoms and 
molecules. In the special 
case when the property of interest is expressed as an expectation value of a 
Hermitian operator 
$\hat{X}$ one considers the following quantity
\begin{align}
\label{deltax}
 \Delta X = \langle \psi | \hat{X} \psi \rangle - \langle \psi_A | \hat{X}_A 
\psi_A \rangle - 
\langle \psi_B | \hat{X}_B \psi_B \rangle,
\end{align}
for a system described by the  wave function $\psi$, and composed of two 
subsystems (monomers 
$A$, $B$) with the wave functions $\psi_A$, $\psi_B$. The operators $\hat{X}_A$ 
and $\hat{X}_B$ are 
defined analogously to $\hat{X}$, but involve summations only over the particles 
belonging
to the subsystems $A$ and $B$, respectively. Note that in the general case 
$\hat{X}\neq\hat{X}_A+\hat{X}_B$. The difference, $\Delta X$, depends on the 
distance, $R$, between the interacting systems and possibly their mutual 
orientations. If the operator $\hat{X}$ is the Hamiltonian of the 
system the quantity $\Delta X$ is called the interaction energy or the 
Born-Oppenheimer (BO) interaction potential. Otherwise, the name 
interaction-induced (or collision-induced) property is   used.

Any interaction-induced property can, in principle, be calculated with help of 
Eq. (\ref{deltax}) 
-- this constitutes the so-called supermolecular approach. In fact, most 
calculations of $\Delta X$
rely on the supermolecular approach since the standard electronic structure 
methods are unable to yield the difference $\Delta X$ directly. A notable 
exception from this rule is the symmetry-adapted perturbation 
theory  (SAPT), see  Refs.~\cite{jeziorski94,sherill11,szalewicz12,jansen14} for 
an extended survey.

 The biggest drawback of the supermolecular method is that it involves a 
significant degree of cancellation  between the terms of Eq. (\ref{deltax}). 
This is especially problematic in weakly interacting 
systems 
where the value of $\Delta X$ can be several orders of magnitude smaller than 
the subtracted 
 terms on the right-hand-side of Eq. (\ref{deltax}).  In practice, $\Delta X$  
is often  smaller
 than the errors  of computing  the individual terms  in   Eq. (\ref{deltax}).  
 
A remedy for this problem is to calculate all terms on the right-hand-side of 
Eq. (\ref{deltax}) in 
a consistent manner, so that these errors cancel  out to a large extent  leaving 
an 
accurate value of $\Delta X$. 
To achieve this, one has to use electronic structure  methods that are size 
consistent, i.e., the 
energies or properties of the system  tend to the correct limit (the sum of 
energies or properties 
of noninteracting  monomers) when the distance   between the subsystems grows to 
infinity 
\cite{bartlett81}. 
The size-consistency requirement is 
critically important and is one of the factors which has led to the success and 
widespread popularity of the coupled-cluster theory, see Ref.~\cite{musial07} 
and references 
therein.

Even if the applied electronic structure model is size consistent, one has to 
face  
a problem stemming from the use of finite basis set expansion of wave functions 
used in  Eq. (\ref{deltax}).
 When the dimer and  monomer  energies are evaluated  using their respective 
basis sets, the dimer energy is artificially 
lowered as the monomers in the dimer calculations have  access to a larger basis 
 set than their own 
basis. It has been recognized a long time ago 
\cite{clementi67,kestner68,jansen69} that this 
artificial lowering, referred to as the basis set superposition error (BSSE), 
cannot be viewed 
as a legitimate part of the interaction energy. In calculations employing 
one-electron basis sets 
(algebraic 
approximation) a prescription for removing the BSSE, called the counterpoise 
(CP) correction, was proposed by Boys and Bernardi \cite{boys70}. It amounts to 
performing 
calculations for the monomers by using the whole dimer  
basis set~\cite{boys70,liu73}. While there 
is still an 
ongoing discussion in the literature about the applicability of this scheme 
\cite{gutowski86,gutowski87,chalbie88,gutowski93,
davidson94,gutowski95,chalbie00,lenthe07,kestner07,baerends14}, especially when 
the 
monomers undergo geometrical deformations \cite{simon96,xantheas96,szalewicz98}, 
when small basis sets are used \cite{liedl98,dunning00}, or when 
basis set extrapolation schemes are employed 
\cite{vanmourik98,halkier99,varandas10}, the CP correction is nowadays 
universally accepted as a default \emph{a posteriori} method for elimination of 
BSSE.

Unfortunately, the situation is different in explicitly correlated methods which 
include all 
interparticle distances directly into trial wave functions. Since these wave 
functions are no longer
composed solely of products of orbitals, it  is not clear how  to define a 
monomer basis set that  would correspond to a given dimer  basis and thus would 
allow a consistent 
dimer and monomer calculations, and an error cancellation.  In other words, in 
explicitly correlated calculations it has not been 
possible thus far   to compute the monomer quantities in Eq. (\ref{deltax}) in 
such a way  that $\Delta X$ 
vanishes in the limit of infinite monomer separations.

In this paper we show how to solve this difficulty. We consider the explicitly 
correlated Gaussian (ECG)
basis  which is arguably the most efficient basis for solving both 
clamped-nuclei and fully non-adiabatic Schr\"{o}dinger equation for few-body 
systems  \cite{bubin13,mitroy13}.
 It has been successfully applied both to light atoms and to small 
molecules, and in many cases the results obtained with ECG are the most accurate 
to date 
\cite{pachucki04,puchalski05,cencek05,cencek08,bubin11,tung11,adamowicz12,
puchalski13,matyus14,przybytek17}. 
by any other method. 
 It should be stressed, however,  that the   method  
proposed by us  can also be applied to calculations with Slater geminals 
\cite{thakkar77,frolov95,korobov02,puchalski10}, Hylleraas CI expansions 
\cite{sims71a,sims71b,sims75}, and other multi-electron basis sets where 
finite-basis size consistency problem arises. 

It should be noted that attempts  to achieve the size consistency of the ECG 
method or to reduce the impact of its violation have  been made and are 
described  in the literature. 
Conceptually the simplest yet practically the most challenging strategy is to 
calculate the 
dimer term in Eq. (\ref{deltax}) as accurately as possible and use the exact or 
near-exact monomer values to get $\Delta X$. This brute-force approach 
typically works well for 
separations where $\Delta X$ is much larger than the error in $\langle \psi | 
\hat{X} \psi 
\rangle$.  However, it does nothing to restore the size-consistency.  As 
$\Delta X$ does not vanish at large $R$ the results deteriorate strongly  with 
increasing $R$ and 
are difficult to match   to an appropriate asymptotic formula. Examples of 
brute-force ECG 
calculations can be found, for example, in Refs.~\cite{rychlewski94,cencek95}.

Another strategy, called the {\em monomer-contraction} (MC) method, has been 
proposed by Cencek \emph{et al.} 
\cite{cencek05,cencek08,patkowski08}.
The main idea  of this method is to build the product of the best available 
monomer wave functions 
into the dimer basis and represent  $\psi$ as 
\begin{align}
 \psi = c_0 \Pi ( \psi_A\psi_B  ) + \sum_k c_k \phi_k,
\end{align}
where $\psi_A$, $\psi_B$ are   wave functions optimized separately for  monomers 
$A$ and $B$, and fixed during the calculations for the dimer, $\Pi$ is a 
projection operator ensuring that $\psi$ has the  correct 
permutation    and spatial symmetry, and $\phi_k$ are elements of the 
conventional ECG basis  for the dimer. The 
rationale behind the MC method is that if the monomer wave functions are 
accurate enough, the
nonlinear optimization of $\phi_k$ is directed mostly towards the 
interaction-induced part of the dimer wave function. 
The monomer quantities entering Eq. (\ref{deltax}) can be computed from 
$\psi_A$, $\psi_B$ or 
more accurate literature values can be used if available. While this 
approach does not fully eliminate the error due to size inconsistency and, 
consequently,  the accuracy breakdown at large $R$, it has been shown to give 
very accurate results for the helium dimer   
in the area of the van der Waals well  \cite{przybytek17}.

A different approach to solve the size-consistency problem in the ECG method 
was 
proposed by Piszczatowski \emph{et al.}  \cite{piszcz08}. In this approach, 
related to SAPT but not 
relying on the convergence 
of a perturbation expansion,  the difference $\Delta X$ is computed 
directly and, by construction, vanishes at large $R$. However, this method  is 
much more computationally expensive than the previous two, as there is a need to 
solve a 
set of response equations for each property of interest. This method has never 
been applied to the 
interaction energy itself.

In the subsequent Sections we shall present our  method to achieve size 
consistency and to eliminate 
BSSE in explicitly correlated calculations, and demonstrate its usefulness for 
the ECG wave 
functions.
Specifically, we shall show how to calculate the  $R$-dependent  sum  of 
monomer 
energies  (or other properties),  corresponding  to a given basis set of 
the dimer, such that $\Delta X$ vanishes at large $R$. 
Therefore, the method can be viewed as a generalization of the conventional CP 
correction 
\cite{boys70} beyond the orbital approximation. 
In fact, our CP correction plays a much more important role than in the orbital 
calculations  
because without it finite basis set explicitly correlated calculations are not 
size 
consistent.  
Taking the interaction of hydrogen atoms as a model system, for which  
practically exact results  are known, we shall demonstrate numerically 
that the proposed technique guarantees  size consistency both in calculations of 
the interaction energy 
and interaction-induced properties.

Atomic units are used throughout the present work unless explicitly stated 
otherwise. We assume that the value of the fine-structure constant, $\alpha$, is $1/137.0359997$.

\section{Theory}
\label{sec:theory}

We assume that the wave functions $\psi$, $\psi_A$, and $\psi_B$, employed in 
Eq. (1) 
to compute $\Delta X$, are approximations to the exact eigenfunctions of the 
electronic Hamiltonians 
$\hat{H}$,   $\hat{H}_A$, and $\hat{H}_B$, and are obtained  using the 
Rayleigh-Ritz variational procedure with  the ECG basis.  
For a diatomic molecule (or a dimer) consisting of atoms with $N_A$ and $N_B$ 
electrons the generic 
ECG  function can be expressed in the form 
\begin{align} 
  \label{phi}
  \phi =
  \prod_{i=1}^{N} e^{ -\alpha_{i} |{\br}_i - {\ba}|^2 }\ 
 \prod_{i=1}^{N} e^{ -\beta_{i} |{\br}_i - {\bb}|^2 } 
  \!\!\prod_{i>j=1}^{N} \!\! e^{ -\gamma_{ij} |{\br}_i - {\br}_j|^2 },
\end{align} 
where  ${\br}_i$, $i=1,\ldots ,N$, are vectors containing   Cartesian 
coordinates
of electrons, $\ba$ and $\bb$ are vectors specifying the nuclear positions, 
$R=|\ba -\bb|$,  and 
$N=N_A+N_B$.  The exponents  $\alpha_i$, $\beta_j$, and $\gamma_{ij}$ are 
different for each basis function, 
and  are optimized by minimizing the lowest eigenvalue of the Hamiltonian 
matrix. 
For simplicity we assumed that the dimer is in a $\Sigma^+$ state.  The 
functions of the form of Eq. 
(\ref{phi})  constitute  a potentially complete basis set in the space  of 
$\Sigma^+$ symmetry  
\cite{jeziorski97,hill08}.  To 
construct   ECGs of other symmetries one can follow the prescription  of 
Ref.~\cite{jeziorski97}.  
The ECG basis functions for the monomer A (B)  can also be  expressed using  Eq. 
(\ref{phi}) 
provided that $N$ is  replaced by  $N_A$ ($N_B$) and the factors 
$e^{ -\beta_{i} |{\br}_i - {\bb}|^2 }$  \ $(e^{ -\alpha_{i} |{\br}_i - {\ba}|^2 
})$
are eliminated.  

We assume that the Hamiltonians  $\hat{H}$,   $\hat{H}_A$, and $\hat{H}_B$ are 
non-relativistic and 
do not act on spin variables. Therefore, we can employ the spin-free formalism 
where the correct 
spin symmetry and fulfillment of the Pauli exclusion principle are 
simultaneously guaranteed by 
imposing the appropriate  permutation symmetry  of the wave 
function~\cite{patkowski01,patkowski02,Kaplan:17}.   Specifically for a system 
with $N$ electrons and spin $S$ the wave function must transform according to 
the  irreducible representation 
of the permutation  group S$_N$  corresponding to the Young diagram containing 
$N/2 -S$ rows of length 
2 and  2$S$ rows of length 1, denoted conventionally as  $[2^{N/2-S}1^{2S}]$.  
This symmetry of the 
wave function can be enforced with the help of  appropriate Young operators 
\cite{Kaplan:17}. Within 
the present computational capabilities this  spin-free ECG method is applicable 
to systems 
containing up to seven/eight active particles, see 
Refs.~\cite{bubin09,sharkey14} as a 
representative examples.

By inspection of Eq. (\ref{phi}) we see that for a given $N$-electron dimer 
basis it is difficult to 
construct the corresponding $N_A$-electron  and $N_B$-electron bases for the 
monomers such that finite basis set calculations will be size consistent. 
Indeed, to the best of our knowledge, no such construction has been proposed in 
the literature. One reason for this difficulty is the inherent delocalization of 
the dimer basis set functions. Another reason is the fact that the basis 
functions used to expand $\psi$ and $\psi_A$ or $\psi_B$ depend on different 
number of electrons. The latter difficulty can be circumvented if Eq. 
(\ref{deltax}) is rewritten in the form 
\begin{align}
\label{2}
 \Delta X = \langle \psi | \hat{X} \psi \rangle - \langle \psi_0 | \hat{X}_0 
\psi_0 \rangle,
\end{align}
where $\hat{X}_0= \hat{X}_A +  \hat{X}_B$,  and $\psi_0=\psi_A \psi_B$ is the 
appropriate 
eigenfunction 
of   $\hat{H}_0= \hat{H}_A +  \hat{H}_B$, i.e., 
\begin{align}
\label{0}
  \hat{H}_0 \psi_0 = (E_A +E_B)\psi_0,
\end{align}
where $E_A=\langle \psi_A | \hat{H}_A \psi_A \rangle$ and 
 $E_B=\langle \psi_B | \hat{H}_B \psi_B \rangle$.   We assume  for simplicity 
that the $E_0$ level of $\hat{H}_0$ 
is non-degenerate.  The functions $\psi$ and $\psi_0$  depend on the same 
number electronic coordinates and thus can, in principle, be obtained 
  by diagonalizing matrices of the Hamiltonians $\hat{H}$ and 
$\hat{H}_0$, respectively, within the same basis set. The calculations performed 
in this way would 
indeed be consistent, so that one could expect both the error cancellations to 
occur and $\Delta 
X$ to correctly vanish at large  separations. 

The problem with this idea is that  $\psi$ and $\psi_0$ have 
different symmetries and, even at  large $R$,  reside in distant  locations of 
the Hilbert space 
\cite{Jeziorski:77}.  This is a consequence of the fact that $\hat{H}$ and 
$\hat{H}_0$ have different symmetry groups, denoted by  $\mathcal{G}$ and 
$\mathcal{G}_0$ further in the text.  
It is impossible to perform calculations for $\psi$ and $\psi_0$  in a common 
basis adapted to 
irreducible representations of both $\mathcal{G}$ and $\mathcal{G}_0$.  The main 
idea of our method 
is to perform calculations with the basis that is adapted to 
$\mathcal{G}\cap\mathcal{G}_0$, i.e.,  
the largest subgroup  of $\mathcal{G}$ and $\mathcal{G}_0$. To guarantee that 
$\psi$ is a pure spin 
state we also impose the condition that this basis is invariant under all 
operations of 
$\mathcal{G}$. 

To illustrate this idea with a simple example we assume that A and B are 
ground-state hydrogen 
atoms.   
In this case $\mathcal{G}=D_{\infty h}\times {\rm S}_2$  and  
$\mathcal{G}_0=O_{\ba}(3)\times O_{\bb}(3)\times G_I$, where $O_{\ba}(3)$ and\ 
$O_{\bb}(3)$ are symmetry groups of $H_A$ and    $H_B$, respectively, and  
$G_I=\{E,P^*\}$ is the two-element group containing the identity element  $E$ 
and the permutation-inversion operation
 $P^* = \hat{I}P_{12}$. The latter is a combination of the  inversion $\hat{I}$ 
with 
respect to the center of the diatom and the transposition   $P_{ij}$ of the 
coordinates of the $i$th 
and $j$th electron.  The groups $O_{\ba}(3)$  and 
$ O_{\bb}(3)$ contain all rotations and the inversion  with respect to the 
respective nuclear positions $\ba$ and $\bb$ (the accidental SO(4) symmetry  of 
hydrogen atom can be neglected as it is not relevant in further discussion).    
 
It is easy to see that the largest common subgroup of $\mathcal{G}$ and 
$\mathcal{G}_0$ is the group $C_{\infty v}\times  G_I$. 
The primitive ECG function of Eq. (\ref{phi}) is already adapted to $C_{\infty 
v}$.  To 
additionally adapt this basis to $G_I$ we project it with $(1+P^*)/2$  (we take 
the plus sign 
in the projector since both
$\psi_0$ and $\psi$ are symmetric under the action of  $P^*$).  The basis 
adapted to  $\mathcal{G} \cap \mathcal{G}_0$ 
consists thus of  functions  of the form
 \begin{align} 
  \label{phi0}
  \begin{split}
   \phi^{\prime}& =
   e^{ -\alpha_{1}  r_{a1}^2  }\ e^{ -\alpha_{2} r_{b1}^2 }\  e^{ -\beta_{1}  
r_{a2}^2  }\ e^{ 
-\beta_{2} r_{b2}^2 }\   e^{ -\gamma_{ij} r_{12}^2 } \\
 & +  e^{ -\beta_{2}  r_{a1}^2  }\ e^{ -\beta_{1} r_{b1}^2 }\  e^{ -\alpha_{2}  
r_{a2}^2  }\ e^{ 
-\alpha_{1} r_{b2}^2 }\   e^{ -\gamma_{ij} r_{12}^2 },
  \end{split}
\end{align} 
where  $r_{ai}=|{\br}_i - {\ba}|$,  $r_{bi}=|{\br}_i - {\bb}|$, and 
$r_{ij}=|{\br}_i - {\br}_j|$.  
This basis is not invariant under the operations of $\mathcal{G}$, so we have to 
augment it by adding  
functions $\hat{I}\phi^{\prime}$ and 
$P_{12}\phi^{\prime}$.  Both augmentations lead to the same result so the final 
basis consists of   
functions of the form of $\phi^{\prime}$ and $P_{12}\phi^{\prime}$.  In Section 
\ref{sec:numeric} 
we shall show that variational Rayleigh-Ritz  calculations employing  this basis
both for the dimer and for the monomer (diagonalizing the $\hat{H}$ and 
$\hat{H}_0$ Hamiltonians, 
respectively, and optimizing nonlinear parameters only at the dimer level) are 
consistent in the 
sense that the monomer errors cancel out and the interaction energy approaches 
zero at infinity.  
In practice is it useful to follow the idea of  the monomer-contraction method  
\cite{cencek05,cencek08,patkowski08} and extend this basis by two additional 
functions: 
${\tilde \psi}_A  {\tilde \psi}_B$ and $P_{12}{\tilde \psi}_A  {\tilde \psi}_B$ 
where ${\tilde \psi}_A$ is the best available ECG approximation of the wave 
function for atom  A and ${\tilde \psi}_B= P^*{\tilde \psi}_A$. These
two basis functions  are fixed and, unlike all functions of the form 
$\phi^{\prime}$ and 
$P_{12}\phi^{\prime}$, are not subject  to  the nonlinear optimization.  

Since the nonlinear optimization performed at the dimer level is very 
time-consuming 
it is useful to adapt the whole basis at this stage of  calculations. This is 
possible since the whole basis  
is invariant under the operations the dimer symmetry group $\mathcal{G}$. If  
one is interested in the triplet $^3\Sigma^+_u$ 
state then the size of the basis can be reduced by the factor of two by taking 
only the functions of the form
$(1-P_{12})\phi^{\prime}$  [plus possibly the single  function 
$(1-P_{12}){\tilde \psi}_A {\tilde \psi}_B$]. These basis functions are 
obviously antisymmetric under $P_{12}$ (are triplet functions) but are also 
{\em ungerade}  under the  action of the inversion operator $\hat{I}$  since    
$\hat{I}\phi^{\prime}= P_{12} \phi^{\prime}$ and, consequently,
\begin{align} \label{u}
  \hat{I}(1-P_{12})\phi^{\prime} = -(1 -\hat{I})\phi^{\prime} = - 
(1-P_{12})\phi^{\prime}. 
\end{align}
It is easy to verify  that Eq. (\ref{u}) holds also when $ \phi^{\prime}$ is 
replaced by ${\tilde \psi}_A  {\tilde \psi}_B$ and that the singlet functions 
obtained by the symmetrization $1+P_{12}$ have {\em gerade} symmetry, i.e., are 
invariant under the inversion $\hat{I}$. It should be emphasized that 
a simple diagonalization of $H_0$ in the space of antisymmetric functions 
$(1-P_{12})\phi^{\prime}$ 
only would lead to a completely wrong  energy $E_0$ since at large $R$ the exact 
function $\psi_0$  has 
  equally 
large components in the spaces of symmetric and antisymmetric functions.  Thus, 
the diagonalization 
of $H_0$ and calculation of $E_0$ must be done in the space containing functions 
of both 
symmetries, i.e.,  in the space containing both $ \phi^{\prime}$  and $P_{12} 
\phi^{\prime}$. 

When the interacting one-electron systems  are different, as in the case of 
He$^+\cdots$H interaction,  
$\mathcal{G}=C_{\infty v}\times {\rm S}_2$,    $\mathcal{G}_0=O_{\ba}(3)\times 
O_{\bb}(3) $, and
$\mathcal{G}\cap\mathcal{G}_0=C_{\infty v}$.
The  inversion  symmetry  is not present and the basis for the monomer 
calculations is constructed from the functions  
$\phi$ and $P_{12}\phi$, where $\phi$ is the two-electron primitive ECG [given 
by the first term on the r.h.s. 
of  Eq. (\ref{phi0})]. In the dimer calculations (involving the optimization of 
the nonlinear 
parameters) the basis is half as large and 
consists of the functions  $(1-P_{12})\phi$ (for the triplet state).

The generalization of this construction to the interaction of many-electron 
atoms is natural but 
technically somewhat complicated due to the multidimensionality of the 
representations of the permutation  group. 
The dimer group $\mathcal{G}$ contains now the factor S$_N$ instead of S$_2$ and 
one has to 
include in $\mathcal{G}_0$ the product S$_{N_A}\times$S$_{N_B}$ of the monomer 
permutation groups.  
Similarly  as for the H$_2$  the basis is constructed in two steps. First, the 
primitive 
ECG basis of Eq. (\ref{phi}) is adapted to the appropriate irreducible 
representation $\Gamma_0$ of 
$\mathcal{G} \cap\mathcal{G}_0$. Next, one forms the basis of the induced 
representation 
 $\Gamma  \!\uparrow \!\mathcal{G} $ and takes the functions adapted  
simultaneously to $\Gamma_0$
and to the irreducible representations of  $\mathcal{G}$ entering 
$\Gamma_0\!\!\uparrow \!\mathcal{G} $. 
Below we shall illustrate this general procedure with three simple but typical 
examples. 
\vspace{1ex}

\noindent {\em Example 1: Interaction of a singlet helium atom with a hydrogen 
atom}
\vspace{0.5ex}

In this case the dimer symmetry  is   $\mathcal{G}=C_{\infty v}\!\times {\rm 
S}_3$,   while
$\mathcal{G}_0=O_{\ba}(3)\times O_{\bb}(3)\times  {\rm S}_2$, and 
$\mathcal{G}\cap\mathcal{G}_0=C_{\infty v}\!\times {\rm S}_2$.  For the singlet 
state of 
helium the function  $\psi_0=\psi_{\rm He}\psi_{\rm H}$ is symmetric 
and the molecular $^2\Sigma^+$ function can be chosen to be symmetric under the 
permutation  
$P_{12}$. 
Therefore, we can symmetrize the  ECG basis and consider further the functions  
$\phi^{\prime} 
=(1+P_{12})\phi$, where
$\phi$ is a primitive, three-electron, two-center ECG function of the form of 
Eq. (\ref{phi}).  To obtain the basis 
invariant under the action of ${\rm S}_3$ we have to perform the induction 
process, i.e., act on $\phi^{\prime}$
with all permutations from  ${\rm S}_3$. In this way we obtain three ECG 
functions   
$\phi^{\prime}$, $P_{13} \phi^{\prime}$, and $P_{23}\phi^{\prime}$ forming a 
basis for the 
induced representation $[2]\uparrow{\rm S}_3$.  The representation  $ 
[2]\uparrow{\rm 
S}_3$, referred also
as the outer product $[2]\otimes[1]$   (see Ref.~\cite{Kaplan:17}),
is reducible and decomposes as  
 \begin{align}
 \label{21x1}
 [2]\uparrow{\rm S}_3 = [21] +[3] . 
\end{align}
which can also be represented with help of the Young diagrams as
\begin{align}
\label{21x1_2}
\begin{ytableau}
 \phantom0 & \phantom0 \\
\end{ytableau}
\;\;\otimes\;\;
\begin{ytableau}
 \phantom0 \\
\end{ytableau}
\;\;=\;\;
\begin{ytableau}
 \phantom0 & \phantom0 \\
 \phantom0 \\
\end{ytableau}
\;\;+\;\;
\begin{ytableau}
 \phantom0 & \phantom0 & \phantom0 \\
\end{ytableau}  \ \ .
\end{align}
One of the two functions transforming according to the [21] representation is 
antisymmetric  under the action of $P_{12}$ and can be disregarded. We are left 
with the functions
\begin{align}
\label{phibis} 
\phi^{\prime\prime} = (2-P_{13}-P_{23}) \phi^{\prime} 
\end{align}
that can be used in  calculations of  the physical, spin doublet state  of the 
molecule, and the 
functions 
\begin{align}
\label{phiter}
 \phi^{\prime\prime\prime} = (1 +P_{13}+P_{23}) \phi^{\prime} 
\end{align}
 that are  Pauli forbidden (cannot be used to construct an antisymmetric 
spin-dependent function)  but must be used 
together with   $\phi^{\prime\prime}$  in consistent  calculation of the sum of 
monomer energies. 
Equation (\ref{phibis}) can be obtained by acting on $\phi^{\prime}$ with  
character projector  of the 
[21] representation of S$_3$, or directly from 
$\phi$ by acting with 
the Young operator  $\omega_{11}^{[21]}$ corresponding to the orthogonal 
Young-Yamanouchi
representation [21] of  S$_3$. In general
  \begin{align}
 \label{gengen}
  \omega_{rt}^{[\lambda]} = \sum_{P \in S_N} \Gamma_{rt}^{[\lambda]}(P) \, P,
 \end{align} 
\vspace{-2ex} 

\noindent where $\Gamma_{rt}^{[\lambda]}(P)$ are matrices of the   
representation $[\lambda]$ \cite{Kaplan:17}. 
 Since the nonlinear parameters are optimized only for the dimer, these 
parameters 
are identical in $\phi^{\prime\prime\prime} $ and $\phi^{\prime\prime} $. 
 Equations 
(\ref{21x1}) and (\ref{phiter}) show that to obtain size consistent energy one 
has to  violate the Pauli principle in calculations of the sum of the monomer 
energies.
\vspace{1ex} 

\noindent {\em Example 2: Interaction of a doublet lithium atom with a hydrogen 
atom}
\vspace{1ex}

This case considered, e.g., in Refs.~\cite{patkowski01,patkowski02}, is 
somewhat more complicated since we have the exchange degeneracy for lithium  
and the S$_4$ group is larger than  S$_3$. The groups $\mathcal{G}$, 
$\mathcal{G}_0$ and 
$\mathcal{G}\cap\mathcal{G}_0$ are the same as in the previous example except 
that the S$_3$
factor in  $\mathcal{G}$ is replaced by S$_4$, and the S$_2$ factor in 
$\mathcal{G}_0$ 
by S$_3$. The doublet states of  lithium exhibit (unphysical) exchange 
degeneracy since the [21] representation is two dimensional  and we have two 
standard Young tableaux 
\begin{align}
\label{S3}
\begin{ytableau}
 1 & 2 \\
 3 \\
\end{ytableau}
\;\;\;\;\;\mbox{and}\;\;\;\;\;
\begin{ytableau}
 1 & 3 \\
 2 \\
\end{ytableau} \,\,  \ .
\end{align}
We  (arbitrarily) chose the first one and require that the lithium wave 
function 
$\psi_{\rm Li}$ 
as well as the molecular  function $\psi_{\rm LiH}$  are symmetric with respect 
of the exchange of 
the electrons $1$ and $2$. Thus, diagonalizations for both the supermolecule  
and the noninteracting  monomers   can be performed in the space 
with the permutational symmetry specified by the first tableau in 
Eq.~(\ref{S3}).  To construct a basis of this symmetry
for   consistent  molecule and separated atom calculations  we start by 
projecting the primitive 
four-electron ECG function of Eq.~(\ref{phi})  with   $\omega_{11}^{[21]}$,  
\begin{align}\label{philih} 
\phi^{\prime} =  (2-P_{13}-P_{23})(1+P_{12}) \phi
\end{align} 
and generate the induced  representation by acting on $\phi^{\prime}$ with all 
S$_4$ 
permutations. Using, e.g., the  Littlewood  theorem for the outer product 
decomposition \cite{Kaplan:17} we find 
\begin{align}
\label{indlih}  
 [21]\uparrow {\rm S}_4 = [31] \ + \  [22] \ + \ [211].
\end{align}
or by using the Young diagrams
\begin{align}
\label{indlih_2}
\begin{ytableau}
 \phantom0 & \phantom0 \\
 \phantom0 \\
\end{ytableau}
\;\;\otimes\;\;
\begin{ytableau}
 \phantom0 \\
\end{ytableau}
\;\;=\;\;
\begin{ytableau}
 \phantom0 & \phantom0 & \phantom0 \\
 \phantom0 \\
\end{ytableau}
\;\;+\;\;
\begin{ytableau}
 \phantom0 & \phantom0 \\
 \phantom0 & \phantom0 \\
\end{ytableau}
\;\;+\;\;
\begin{ytableau}
 \phantom0 & \phantom0 \\
 \phantom0 \\
 \phantom0 \\
\end{ytableau} \ \ .
\end{align}
The dimension of  the  $[21]\!\uparrow {\rm S}_4$ representation is $8$ but  by 
inspecting the standard Young tableaux we find that there are only three 
functions 
of  the  S$_3$  symmetry corresponding to the first tableau of Eq. (\ref{S3}). 
These three
functions can be obtained directly from  $\phi$  using the  Young operators 
$\omega_{22}^{[31]}$ , 
$\omega_{11}^{[22]}$, and $\omega_{11}^{[211]}$ or
by acting with the [31], [22], and [211] character projectors  on 
$\phi^{\prime}$. 
The explicit form of these three functions is  
\begin{align}
\label{31}
\phi^{[31]} & =  \big(3+P_{34}\big)\big(1+P_{14}+P_{24}\big)\phi^{\prime}, \\ 
\label{22}
\phi^{[22]} & =  \big(1+P_{34}\big)\big(1-P_{14}-P_{24}\big)\phi^{\prime}, 
\\\label{211}
\phi^{[211]}&=  \big(1-P_{34}\big)\big(3-P_{14}-P_{24}\big)\phi^{\prime}.
\end{align} 
If we are interested in the  singlet or the triplet states we use the functions 
of the form of $\phi^{[22]}$ or  $\phi^{[211]}$, respectively, while for the 
monomer 
energy calculation we must use (without further nonlinear optimization) both of 
these
functions  plus  the Pauli forbidden one, $\phi^{[31]}$. Thus the basis of the
monomer calculations is three times as large as in the dimer case. 
 \vspace{1ex} 

\noindent {\em Example 3: The ground state of the helium dimer}
\vspace{1ex}

In this case we have to consider both the permutation and the inversion 
symmetry. The groups 
$\mathcal{G}$, $\mathcal{G}_0$ and $\mathcal{G}\cap\mathcal{G}_0$ have now the 
following direct product structure: 
$\mathcal{G}= D_{\infty h} \times {\rm S}_4$, \ 
$\mathcal{G}_0=O_{\ba}(3)\!\times\! O_{\bb}(3)\times  {\rm S}_2\!\times\!  {\rm 
S}_2 \!\times\! \mathcal{G}_I$, and 
$\mathcal{G}\cap\mathcal{G}_0={\rm C}_{\infty v}\!\times\!  {\rm S}_2\!\times\!  
{\rm S}_2 \!\times\!  \mathcal{G}_I$,
 where $\mathcal{G}_I = \{E, \hat{I} P_{ab}\}$ is a two-element group containing 
the product of the inversion 
operation $\hat{I}$ and a permutation $P_{ab}$ that swaps all electrons between 
atoms A and B. Assuming that $H_A$ and $H_B$ act on electrons 1,2 and 3,4, 
respectively, the permutation $P_{ab}$ can be taken arbitrarily as any of the 
four $P_{13}P_{24}$, $P_{14}P_{23}$, $P_{1324}$, or $P_{1423}$ without changing 
the $ {\rm S}_2\!\times\!  {\rm S}_2 \!\times\!  \mathcal{G}_I $ group. 

Since we are interested in the interaction of singlet states we can adapt the 
ECG basis to the 
fully symmetric representation of $\mathcal{G}\cap\mathcal{G}_0$ and use the 
symmetrized ECG 
functions of the form
\begin{align} \label{phi1he2}
\phi^{\prime} = (1+P_{12})(1+P_{34})(1+\hat{I} P_{ab})\phi,
\end{align} 
where $\phi$ is the primitive ECG function of Eq. (\ref{phi}) with $N=4$. 
Performing the induction 
of the fully symmetric representation of  
${\rm S}_2\!\times\!  {\rm S}_2$ to S$_4$ one finds
\begin{align} \label{indhe2}
[2]\!\times\![2]\! \uparrow \!{\rm S}_4  = [4] + [31] + [22].
\end{align}
which can be written by using the Young diagrams as
\begin{align}
\label{indhe2_2}
\begin{ytableau}
 \phantom0 & \phantom0 \\
\end{ytableau}
\;\otimes\;
\begin{ytableau}
 \phantom0 & \phantom0 \\
\end{ytableau}
\;=\;
\begin{ytableau}
 \phantom0 & \phantom0 & \phantom0 & \phantom0 \\
\end{ytableau}
\;+\;
\begin{ytableau}
 \phantom0 & \phantom0 & \phantom0 \\
 \phantom0 \\
\end{ytableau}
\;+\;
\begin{ytableau}
 \phantom0 & \phantom0 \\
 \phantom0 & \phantom0 \\
\end{ytableau} \   .
\end{align}
The induction from ${\rm C}_{\infty v}$ to $D_{\infty h}$ is not needed since 
one can show that 
the space spanned by  $P\phi^{\prime}$, $P \in {\rm S}_4$ is invariant under the 
inversion $\hat{I}$. 
The induced representation $[2]\!\times\![2]\! \uparrow \!{\rm S}_4  $  is 
six-dimensional but, since we can
work only with functions fully symmetric under ${\rm S}_2\!\times\!  {\rm S}_2$, 
only three 
functions are necessary
\begin{align}
\label{he2ecgs} 
\phi^{[4]}  &= \big(1+  P_{13}\,P_{24} + P_{13}+ P_{14}+  P_{23}+ 
P_{24}\big)\phi^{\prime},\\
\phi^{[31]} &= \big(1-P_{13}P_{24}\big)\phi^{\prime},\\
\phi^{[22]} &= \big(2+ 2\,P_{13}P_{24} - P_{13} - P_{14} - P_{23} - 
P_{24}\big)\phi^{\prime}.
\end{align}
Only the last of these functions is Pauli allowed and appears in the dimer 
calculations. The first 
two are Pauli forbidden but must be used in calculations for the noninteracting 
monomers  to obtain 
size-consistent results.

The functions $\phi^{[22]}$ and $\phi^{[4]}$ are already of the {\em gerade} 
symmetry under inversion. To prove this we note that  
\begin{align}\label{p1}
\hat{I} \phi^{\prime}  = P_{ab}\,\phi^{\prime},
\end{align}
for any of the  four permutations $P_{ab} $. One can show that the parts of 
$\phi^{[4]}$ or $\phi^{[31]}$ generated by 
$1+ P_{13}\,P_{24}$, by $ P_{13}+P_{24}$ and by $P_{23}+P_{14}$ are separately 
invariant under the action of $\hat{I}$. Specifically  
\begin{align}\nonumber
\hat{I}(P_{13}\!+\!P_{24}) \phi^{\prime}\! =\!
\hat{I}\phi^{\prime}=  
(P_{13}\!+\!P_{24}) P_{13} P_{24}\phi^{\prime} \! =\! 
(P_{13}\!+\!P_{24})\phi^{\prime},\\
\nonumber 
\hat{I}(P_{23}\! +\!P_{14}) \phi^{\prime}\! =\!
\hat{I}\phi^{\prime}=  
(P_{23}\!+\!P_{14}) P_{23} P_{14}\phi^{\prime} \! =\! 
(P_{23}\!+\!P_{14})\phi^{\prime},
\end{align}
and similarly for $(1+P_{13}P_{24})\phi^{\prime}$. Analogously one can show 
that 
$\phi^{[31]}$ is {\em ungerade}  under inversion.  
 
In all examples considered here the induced representation is simply reducible. 
However, there are 
cases when there are multiplicities. For instance, for the interaction of three 
ground-state hydrogen 
atoms, the representation $[21]$ occurs two times. Therefore, in the trimer 
calculations we have two ECG functions of the $[21]$ symmetry for one primitive 
ECG function. To 
obtain the energy of the monomers all six ECG functions  spanning the induced 
(regular in this case) representation must be used. Similar multiplicity problem 
occurs for the 
interaction of two doublet lithium atoms  when the  representation $[321]$ 
appears two times in the 
direct product $[21]\otimes[21]$.

\section{Numerical results}
\label{sec:numeric}

\subsection{Computational details}

\begin{table}
\caption{\label{tab:mc}
Properties of the hydrogen atom calculated with the orbital 1$s$ expanded in 9 
or 12 primitive 
Gaussian functions.  
$E$ is the electronic energy and $\langle\delta(\textbf{r})\rangle$ is the 
expectation value 
of   delta distribution centered at the nucleus.  Errors with respect to the 
exact values are given below each 
entry.}
\begin{ruledtabular}
\begin{tabular}{cccc}
  & 9 & 12 \\
 \hline\\[-2.2ex]
$E$   & $-0.499\,998\,136$ & $-0.499\,999\,904$ \\
error & $\phantom{+}0.000\,001\,864$ & $\phantom{+}0.000\,000\,096$ \\
\hline\\[-2.2ex]
$\langle\delta(\textbf{r})\rangle$ & $\phantom{+}0.317\,799\,920$ & 
$\phantom{+}0.317\,840\,649$ \\
error                              & $\phantom{+}0.000\,509\,966$ & 
$\phantom{+}0.000\,469\,238$ \\
\end{tabular}
\end{ruledtabular}
\end{table}

As a numerical illustration we performed  variational ECG calculations of the 
interaction energy of hydrogen atoms in the ground ($^1\Sigma_g^+$) state of 
H$_2$. We employed the monomer contraction   method of Cencek 
\emph{et al.} and assumed  the trial wave function   in the  form
\begin{align}
\label{ecg0}
 \left(1+P_{12}\right)\left(1+P_{ab}\right) \bigg[ 
c_0\,\phi_{1s}(r_{1a})\,\phi_{1s}(r_{2b})+\sum_{k=1}^K c_k 
\phi_k \bigg],
\end{align}
where $\phi_k$ are the primitive geminal functions, cf. Eq. (\ref{phi}) with 
$N=2$,
 \begin{align}
\label{ecgh2}
 \phi_k = e^{-a_k r_{1a}^2 - b_k r_{1b}^2 - c_k r_{2a}^2 - d_k r_{2b}^2 - w_k 
r_{12}^2},
\end{align}
and $\phi_{1s}(r)$ is the hydrogenic 1$s$ orbital expanded as a linear 
combination 
of Gaussian $1s$ functions. Two distinct basis sets were optimized -- 
the first composed of 150 geminal functions with the monomer contraction length 
of nine functions ($9/150$ basis set) and the second composed of 300 geminal 
functions with the monomer contraction length of twelve functions ($12/300$  
basis set).
The relevant  properties of the hydrogen atom obtained for each monomer 
contraction function 
are given in Table~\ref{tab:mc}.

The non-linear parameters $a_k$, $b_k$, \emph{etc.} in all functions 
(\ref{ecgh2}) were optimized 
to minimize the total energy of the molecule. We employed the conventional 
optimization 
strategy where the primitive functions are optimized one at a time using the 
Powell's conjugate 
direction method \cite{powell64}. Technical details of this procedure can be 
found, for example, in 
Refs.~\cite{komasa93,komasa01}. About one thousand optimization sweeps over the 
whole basis set were 
performed for each internuclear distance. The monomer contraction functions were 
kept fixed during 
the optimization procedure. The distance-dependent  energies of noninteracting 
monomers were obtained 
according to the prescription given in the previous section.

In the present case the monomers are one-electron atoms and thus it 
would theoretically be possible to use even more accurate monomer contraction 
functions, i.e., accurate down to the level of the arithmetic precision. 
However, our goal here is to simulate the situation found in other systems, 
e.g., the helium dimer, where such accurate monomer contractions are 
practically unfeasible.

\subsection{Interaction energies}

The simplest numerical confirmation of the size consistency of  the proposed 
counterpoise correction  can be obtained by applying it 
to compute the interaction energy with only a single ECG basis set function. 
This test can be viewed as the most demanding one  as it is 
well-known that the size-consistency problems are much more pronounced in 
smaller basis sets. For the purposes of this test we did not use the MC method.
  The single ECG basis   function was optimized separately for each $R$ to get 
the best possible energy of the molecule.

The results of the test for the hydrogen molecule are reported in Table 
\ref{esing} and demonstrate 
that the dimer energy and the energy of noninteracting monomers  tend to the 
same value for large $R$. 
Thus, the interaction energy vanishes at large $R$ and one obtains 
size-consistent results. It is of note 
that for  $R>7.0$ the energy of the noninteracting monomers becomes practically 
independent of $R$.  We performed a similar test also  
  for the HeH molecule in the ground 
($^2\Sigma^+$) state, see Table~\ref{esing2}. This provides a   verification 
that the proposed 
counterpoise correction works for a three-electron system with a nontrivial 
permutation symmetry.
During the  optimizations for the HeH molecule we frequently encountered 
multiple local minima and had to pay  attention to avoid jumping between them 
when the internuclear distance was increased. 
We checked that after applying the counterpoise correction the interaction 
energy 
vanished at large $R$, independently of which local minimum was selected in the 
calculations.

\begin{table}
\caption{Dimer energy and the counterpoise corrected energy of noninteracting 
atoms for the hydrogen molecule 
(H$_2$) in the $^1\Sigma_g^+$ state calculated with a single ECG function. The  
difference 
between the two energies is given in the last column.   The symbol $X_{\pm n}$ 
stands for 
$X\cdot 10^{\pm n}$.}
\begin{ruledtabular}
\begin{tabular}{cccl}
\label{esing}
$R$ & \multicolumn{1}{c}{dimer energy} & \multicolumn{1}{c}{monomer energies} & 
\multicolumn{1}{c}{diff.} \\
\hline\\[-2.2ex]
1.40 & $-$1.080\,150\,157 & $-$0.851\,504\,752 & 2.29$_{-1}$ \\ 
2.00 & $-$1.047\,848\,806 & $-$0.877\,907\,811 & 1.70$_{-1}$ \\ 
3.00 & $-$0.962\,272\,248 & $-$0.892\,953\,363 & 6.93$_{-2}$ \\ 
4.00 & $-$0.916\,883\,089 & $-$0.902\,594\,831 & 1.43$_{-2}$ \\ 
5.00 & $-$0.906\,403\,817 & $-$0.904\,962\,697 & 1.44$_{-3}$ \\ 
6.00 & $-$0.905\,161\,164 & $-$0.905\,046\,809 & 1.14$_{-4}$ \\ 
7.00 & $-$0.905\,054\,674 & $-$0.905\,048\,043 & 6.63$_{-6}$ \\ 
8.00 & $-$0.905\,048\,301 & $-$0.905\,048\,052 & 2.50$_{-7}$ \\ 
9.00 & $-$0.905\,048\,057 & $-$0.905\,048\,052 & 5.79$_{-9}$ \\ 
10.0 & $-$0.905\,048\,052 & $-$0.905\,048\,052 & 8.17$_{-11}$ \\ 
\end{tabular}
\end{ruledtabular}
\end{table}

Let us now discuss calculations with a larger number of ECG functions.
In Table \ref{intabs} we present absolute errors in the interaction 
energy of H$_2$ obtained using the $9/150$ and $12/300$ basis sets. The 
reference values used in both tables are taken from the work of Pachucki 
\cite{pachucki10} and can be considered exact for the present purposes. 
For each internuclear distance the interaction energy was calculated
employing  the same total dimer energy and  by subtracting:
\begin{itemize}
 \item [$(a)$]  
monomer~energies~calculated~from~the~MC~function~alone~(pure~MC~method);
 \item [$(b)$] exact monomer energies  (exact 
monomer method);
 \item  [$(c)$]  the counterpoise-corrected energy of the noninteracting 
monomers calculated according to the scheme given in Section \ref{sec:theory} 
(CP method);
 \item  [$(d)$]  the large-$R$ asymptotic energy of the noninteracting 
monomers;  in practice, the energy of noninteracting monomers 
computed at the largest available interatomic distance  (asymptotic CP method). 
 \end{itemize}

 Table \ref{intabs} presents 
  results near the minimum of the potential energy curve 
($R=1.4$) and in the long-range tail of the potential.  
It is clearly seen that the methods based 
on subtracting the exact or MC monomer 
energies are not size-consistent as the interaction energies calculated 
with these methods 
tend to some spurious  non-zero values. 
This is best visible for the $9/150$ basis set even for quite moderate $R$, 
whereas for 
the larger $12/300$ basis set the deterioration of the results is less 
pronounced.
 In contrast, the counterpoise-corrected 
interaction energy vanishes as $R\rightarrow\infty$. 

\newcolumntype{d}[1]{D{.}{.}{#1}}
\setlength{\tabcolsep}{-2pt} 

 \begin{table}
\caption{Molecule energy and the counterpoise corrected energy of 
noninteracting 
atoms for the HeH molecule 
 in the lowest $^2\Sigma^+$ state calculated with a single ECG function. The  
difference 
between the two energies is given in the last column.  The symbol $X_{\pm n}$ 
stands for 
$X\cdot 10^{\pm n}$.}
\begin{ruledtabular}
\begin{tabular}{cccl}
\label{esing2}
$R$ & \multicolumn{1}{c}{molecule energy} & \multicolumn{1}{c}{ monomer 
energies} & 
\multicolumn{1}{c}{diff.\;\;\;\;} \\
\hline\\[-2.2ex]
3.00 & $-$2.761\,101\,011 & $-$2.757\,017\,204 & 4.08$_{-3}$ \\
3.50 & $-$2.755\,780\,617 & $-$2.754\,959\,925 & 8.20$_{-4}$ \\
4.00 & $-$2.753\,543\,972 & $-$2.753\,404\,364 & 1.40$_{-4}$ \\
5.00 & $-$2.751\,558\,211 & $-$2.751\,555\,878 & 2.33$_{-6}$ \\
6.00 & $-$2.750\,556\,963 & $-$2.750\,556\,942 & 2.13$_{-8}$  \\
7.00 & $-$2.749\,956\,748 & $-$2.749\,956\,747 & 1.03$_{-9}$  \\
8.00 & $-$2.749\,568\,092 & $-$2.749\,568\,092 & 3.47$_{-10}$ \\
\end{tabular}
\end{ruledtabular}
\end{table}

It is obvious that the relative errors in the interaction energies computed 
using the MC  method 
or the exact monomer energies must grow to infinity at large $R$. 
Inspecting the results of Table~\ref{intabs} one can find, however,  that the 
relative error
 in  both CP approaches also grows with $R$ although   moderately (not 
increasing to infinity at large $R$). 
 This seems discouraging but we 
believe that this  is unavoidable in methods where the wave function is 
optimized variationally at each $R$.  For large $R$ the monomer 
energies constitute the dominating contribution to the total energy of the 
supermolecule and thus the optimization is biased towards improving the monomer 
energies rather than describing the bonding.  This is not a serious problem  in 
practice since at very large $R$
the interaction energy can be accurately represented by its asymptotic expansion 
based 
on monomer properties only (e.g., dynamic polarizabilities). 

The main purpose  of the present work was  to correct the interaction energies 
in 
the  long-range region, i.e., where the size-consistency errors are the most 
pronounced, but it is equally important to validate the proposed strategy for 
smaller internuclear separations. Rather surprisingly, the asymptotic CP turns 
out to be 
superior to other methods near the bottom of the potential energy curve and is
capable of recovering at least one additional significant digit as compared 
with 
the CP technique. The method based on subtracting the exact 
monomer energies  also  gives at small $R$ more accurate results than the CP 
method. 
This situation changes for larger $R$.   When the 
$12/300$ basis set is used  the relative error in the interaction energies 
calculated 
with the CP  method becomes smaller already at $R \approx 12.0$.  
For large internuclear distances the standard and asymptotic CP 
methods give, on average, results of a comparable quality.
In view of 	it its very good   behavior at small $R$ 
 the asymptotic CP appears to be  the 
most promising method for calculation of accurate potential energy surfaces for 
larger systems. While this method introduces a degree of arbitrariness, i.e., 
the choice of a internuclear distance used to evaluate the noninteracting 
monomer energies, it seems to be a pragmatic way to extract the best 
possible results out of a given dimer wave function.

\begin{table*}[t]
\caption{Absolute errors in the interaction energy of the hydrogen molecule 
($^1\Sigma_g^+$ state) as a function of the internuclear distance ($R$). The 
symbol $X_{\pm n}$ stands for $X\cdot 10^{\pm n}$.}
\begin{ruledtabular}
\begin{tabular}{d{-1}d{-1}d{-1}d{-1}d{-1}d{-1}d{-1}d{-1}d{-1}d{-1}d{-1}d{-1}c}
\label{intabs}
  & \multicolumn{4}{c}{$9/150$ }  
    & \multicolumn{4}{c}{$12/300$} \\[-0.7ex]
\;\;\;\;$R$ & 
\multicolumn{4}{c}{
--------------------------------------------------------------}  
    & \multicolumn{4}{c}{\ \ \   
------------------------------------------------------------------}& 
\multicolumn{1}{c}{ \;\;\;\;\;\;\;\;\;Ref.~\cite{pachucki10}}\\
     & \multicolumn{1}{c}{\;\;\;MC} & \multicolumn{1}{c}{\;\;\;\;exact mon.} & 
\multicolumn{1}{c}{\;\;\;CP} & \multicolumn{1}{c}{\;\;\;\;\; asym. CP}  &
\multicolumn{1}{c}{\;\;\;\;\;MC} & \multicolumn{1}{c}{\;\;\;\;exact mon.} & 
\multicolumn{1}{c}{\;\;\;CP} & \multicolumn{1}{c}{\;\;\;\; asym. CP} & 
 \\
\hline\\[-2.2ex]
  1.0  & -3.62\mbox{$_{-6}$} & 1.10\mbox{$_{-7}$} & -2.05\mbox{$_{-6}$} & 
  8.65\mbox{$_{-8}$}  &     -1.67\mbox{$_{-7}$}       &   2.58\mbox{$_{-8}$}    
 
 & 
  -5.13\mbox{$_{-8}$} & 2.16\mbox{$_{-8}$} &    -1.24540\mbox{$_{-1}$}   \\ 
1.4  & -3.61\mbox{$_{-6}$} & 1.14\mbox{$_{-7}$} & -2.07\mbox{$_{-6}$} & 
9.02\mbox{$_{-8}$} & -1.70\mbox{$_{-7}$} & 2.25\mbox{$_{-8}$} & 
-7.57\mbox{$_{-8}$} & 1.83\mbox{$_{-8}$} & -1.74476\mbox{$_{-1}$} \\ 
2.0  & -3.62\mbox{$_{-6}$} & 1.12\mbox{$_{-7}$} & -2.35\mbox{$_{-6}$} & 
8.82\mbox{$_{-8}$} & -1.72\mbox{$_{-7}$} & 2.04\mbox{$_{-8}$} & 
-9.53\mbox{$_{-8}$} & 1.62\mbox{$_{-8}$} & -1.38133\mbox{$_{-1}$} \\ 
4.0  & -3.59\mbox{$_{-6}$} & 1.34\mbox{$_{-7}$} & -1.15\mbox{$_{-6}$} & 
1.11\mbox{$_{-7}$} & -1.51\mbox{$_{-7}$} & 4.17\mbox{$_{-8}$} & 
-1.54\mbox{$_{-8}$} & 3.75\mbox{$_{-8}$} & -1.63903\mbox{$_{-2}$} \\ 
6.0  & -3.69\mbox{$_{-6}$} & 3.42\mbox{$_{-8}$} & -4.87\mbox{$_{-7}$} & 
1.09\mbox{$_{-8}$} & -1.77\mbox{$_{-7}$} & 1.57\mbox{$_{-8}$} & 
-1.75\mbox{$_{-8}$} & 1.15\mbox{$_{-8}$} & -8.35708\mbox{$_{-4}$} \\ 
7.0  & -3.70\mbox{$_{-6}$} & 2.67\mbox{$_{-8}$} & -2.07\mbox{$_{-7}$} & 
3.43\mbox{$_{-9}$} & -1.84\mbox{$_{-7}$} & 7.95\mbox{$_{-9}$} & 
-1.11\mbox{$_{-8}$} & 3.77\mbox{$_{-9}$} & -1.97914\mbox{$_{-4}$} \\ 
8.0  & -3.71\mbox{$_{-6}$} & 1.61\mbox{$_{-8}$} & -6.15\mbox{$_{-8}$} & 
-7.19\mbox{$_{-9}$} & -1.88\mbox{$_{-7}$} & 4.65\mbox{$_{-9}$} & 
-8.84\mbox{$_{-9}$} & 4.74\mbox{$_{-10}$} & -5.56050\mbox{$_{-5}$} \\ 
9.0  & -3.71\mbox{$_{-6}$} & 1.61\mbox{$_{-8}$} & -9.11\mbox{$_{-8}$} & 
-7.18\mbox{$_{-9}$} & -1.89\mbox{$_{-7}$} & 3.40\mbox{$_{-9}$} & 
-2.40\mbox{$_{-9}$} & -7.79\mbox{$_{-10}$} & -1.97818\mbox{$_{-5}$} \\ 
10.0 & -3.71\mbox{$_{-6}$} & 1.47\mbox{$_{-8}$} & -5.80\mbox{$_{-8}$} & 
-8.66\mbox{$_{-9}$} & -1.89\mbox{$_{-7}$} & 3.19\mbox{$_{-9}$} & 
-4.25\mbox{$_{-9}$} & -9.91\mbox{$_{-10}$} & -8.75575\mbox{$_{-6}$} \\ 
11.0 & -3.72\mbox{$_{-6}$} & 1.11\mbox{$_{-8}$} & -4.75\mbox{$_{-8}$} & 
-1.22\mbox{$_{-8}$} & -1.90\mbox{$_{-7}$} & 2.42\mbox{$_{-9}$} & 
-3.52\mbox{$_{-9}$} & -1.76\mbox{$_{-9}$} & -4.50599\mbox{$_{-6}$} \\ 
12.0 & -3.72\mbox{$_{-6}$} & 1.31\mbox{$_{-8}$} & -3.32\mbox{$_{-8}$} & 
-1.02\mbox{$_{-8}$} & -1.90\mbox{$_{-7}$} & 2.32\mbox{$_{-9}$} & 
-1.65\mbox{$_{-9}$} & -1.86\mbox{$_{-9}$} & -2.54597\mbox{$_{-6}$} \\ 
14.0 & -3.71\mbox{$_{-6}$} & 1.32\mbox{$_{-8}$} & -2.43\mbox{$_{-8}$} & 
-1.01\mbox{$_{-8}$} & -1.90\mbox{$_{-7}$} & 2.67\mbox{$_{-9}$} & 
-1.50\mbox{$_{-9}$} & -1.51\mbox{$_{-9}$} & -9.60681\mbox{$_{-7}$} \\ 
16.0 & -3.71\mbox{$_{-6}$} & 1.45\mbox{$_{-8}$} & -1.36\mbox{$_{-8}$} & 
-8.84\mbox{$_{-9}$} & -1.90\mbox{$_{-7}$} & 2.62\mbox{$_{-9}$} & 
-1.34\mbox{$_{-9}$} & -1.56\mbox{$_{-9}$} & -4.19586\mbox{$_{-7}$} \\ 
18.0 & -3.71\mbox{$_{-6}$} & 1.45\mbox{$_{-8}$} & -8.81\mbox{$_{-9}$} & 
-8.81\mbox{$_{-9}$} & -1.89\mbox{$_{-7}$} & 3.38\mbox{$_{-9}$} & 
-8.01\mbox{$_{-10}$} & -8.01\mbox{$_{-10}$} & -2.03341\mbox{$_{-7}$} \\ 
\end{tabular}
\end{ruledtabular}
\end{table*}

The sum of noninteracting monomer energies obtained with the counterpoise 
method 
is distance-dependent -- similarly as in   the standard Boys-Bernardi scheme. 
Therefore, it is 
interesting to investigate what is the dependence of this quantity on $R$. In 
Table \ref{emono} we 
show results obtained with $9/150$ and $12/300$ basis sets. Near the bottom of 
the potential 
energy curve the  energy of noninteracting monomers is  only slightly more 
accurate than the  
energy  corresponding to  the monomer contraction function. For example, with 
the $12/300$ basis set 
at $R=1.40$ the errors of these energies are  $98\,$nH  and $192\,$nH, 
respectively.
In the region $R=4.0-10.0$ the former  error drops sharply by one to two 
orders of magnitude. This 
is a manifestation of the fact that for larger 
internuclear distances the optimization is biased towards the monomer. 
Finally, for $R>12.0$ this error decays slowly or 
fluctuates around some constant value amounting to about $30.0\,$nH and 
$4.0\,$nH for the $9/150$ 
and $12/300$ basis sets, respectively. These fluctuations   are at a low 0.1 nH 
level and are 
artifacts of the non-linear optimization procedure. 

\begin{table}[b]
\caption{Absolute errors in the   energy of two noninteracting hydrogen 
atoms (with respect to the exact value of $-1$) calculated with the 
counterpoise 
method as a function of the internuclear distance.  
 $X_{\pm n}$ stands for $X\cdot 10^{\pm n}$.}
\begin{ruledtabular}
\begin{tabular}{cccccccc}
\label{emono}
$R$ & $9/150$ & $12/300$ \\
\hline\\[-2.2ex]
1.00 & 2.163$_{-6}$ & 7.712$_{-8}$ \\
1.40 & 2.181$_{-6}$ & 9.815$_{-8}$ \\
2.00 & 2.465$_{-6}$ & 1.158$_{-7}$ \\
4.00 & 1.279$_{-6}$ & 5.708$_{-8}$ \\
6.00 & 5.209$_{-7}$ & 3.324$_{-8}$ \\
7.00 & 2.332$_{-7}$ & 1.902$_{-8}$ \\
8.00 & 7.761$_{-8}$ & 1.350$_{-8}$ \\
9.00 & 1.072$_{-7}$ & 5.799$_{-9}$ \\
10.0 & 7.267$_{-8}$ & 7.439$_{-9}$ \\
11.0 & 5.859$_{-8}$ & 5.947$_{-9}$ \\
12.0 & 4.634$_{-8}$ & 3.972$_{-9}$ \\
14.0 & 3.752$_{-8}$ & 4.166$_{-9}$ \\
16.0 & 2.804$_{-8}$ & 3.956$_{-9}$ \\
18.0 & 2.331$_{-8}$ & 4.181$_{-9}$ \\
\end{tabular}
\end{ruledtabular}
\end{table}

\subsection{Interaction-induced properties}

\begin{table*}[t]
\caption{Absolute errors in the interaction-induced one-electron Darwin 
correction to the interaction energy of the 
hydrogen molecule in the $^1\Sigma_g^+$ state calculated with the $9/150$ and 
$12/300$ basis sets. The symbol $X_{\pm n}$ stands for $X\cdot 10^{\pm n}$.}
\begin{ruledtabular}
\begin{tabular}{d{-1}d{-1}d{-1}d{-1}d{-1}d{-1}d{-1}d{-1}d{-1}d{-1}d{-1}d{-1}c}
\label{d1abs}
  & \multicolumn{3}{c}{$9/150$}         
    & \multicolumn{3}{c}{$12/300$} \\[-0.5ex]
\;\;\;\;$R$ & 
\multicolumn{3}{c}{------------------------------------------------------}  
    & \multicolumn{3}{c}{\ \ \      
---------------------------------------------------}& 
\multicolumn{1}{c}{ \;\;\;\;\;\;\;\;\;Refs.~\cite{puchalski17,piszcz08}}\\
     & \multicolumn{1}{c}{\;\;\;\; MC} & \multicolumn{1}{c}{\;\;\;\;  exact 
mon.} & 
\multicolumn{1}{c}{\;\;\;CP} &
\multicolumn{1}{c}{\;\;\;MC} & \multicolumn{1}{c}{\;\;\;\; exact mon.} & 
\multicolumn{1}{c}{\;\;\; CP.} & 
\multicolumn{1}{c}{\;\;\;\;\;\;\;\;} \\
\hline\\[-2.2ex]
1.00 & 5.16\mbox{$_{-3}$} & -8.92\mbox{$_{-4}$} & -3.56\mbox{$_{-4}$} &
1.26\mbox{$_{-3}$} & -9.17\mbox{$_{-4}$} & -3.53\mbox{$_{-4}$} &
3.25779\mbox{$_{-1}$} \\
2.00 & 5.52\mbox{$_{-3}$} & -5.30\mbox{$_{-4}$} & -1.21\mbox{$_{-4}$} &
1.65\mbox{$_{-3}$} & -5.20\mbox{$_{-4}$} & -2.25\mbox{$_{-5}$} &
2.20827\mbox{$_{-2}$} \\
4.00 & 5.57\mbox{$_{-3}$} & -4.83\mbox{$_{-4}$} & -2.00\mbox{$_{-5}$} &
1.74\mbox{$_{-3}$} & -4.33\mbox{$_{-4}$} & 4.29\mbox{$_{-5}$} &
-1.83425\mbox{$_{-2}$} \\
6.00 & 5.54\mbox{$_{-3}$} & -5.09\mbox{$_{-4}$} & -2.43\mbox{$_{-6}$} &
1.75\mbox{$_{-3}$} & -4.23\mbox{$_{-4}$} & 3.72\mbox{$_{-6}$} &
-1.64743\mbox{$_{-3}$} \\
7.00 & 5.54\mbox{$_{-3}$} & -5.16\mbox{$_{-4}$} & -8.85\mbox{$_{-8}$} &
1.72\mbox{$_{-3}$} & -4.53\mbox{$_{-4}$} & 3.80\mbox{$_{-6}$} &
-4.32567\mbox{$_{-4}$} \\
8.00 & 5.54\mbox{$_{-3}$} & -5.09\mbox{$_{-4}$} & -5.18\mbox{$_{-7}$} &
1.72\mbox{$_{-3}$} & -4.53\mbox{$_{-4}$} & 1.51\mbox{$_{-6}$} &
-1.23034\mbox{$_{-4}$} \\
9.00 & 5.54\mbox{$_{-3}$} & -5.11\mbox{$_{-4}$} & -2.65\mbox{$_{-6}$} &
1.70\mbox{$_{-3}$} & -4.69\mbox{$_{-4}$} & 3.70\mbox{$_{-7}$} &
-4.13107\mbox{$_{-5}$} \\
10.0 & 5.54\mbox{$_{-3}$} & -5.12\mbox{$_{-4}$} & -1.70\mbox{$_{-6}$} &
1.71\mbox{$_{-3}$} & -4.60\mbox{$_{-4}$} & -4.28\mbox{$_{-7}$} &
-1.69241\mbox{$_{-5}$} \\
11.0 & 5.54\mbox{$_{-3}$} & -5.12\mbox{$_{-4}$} & -1.65\mbox{$_{-6}$} &
1.70\mbox{$_{-3}$} & -4.70\mbox{$_{-4}$} & -4.10\mbox{$_{-7}$} &
-8.30538\mbox{$_{-6}$} \\
12.0 & 5.54\mbox{$_{-3}$} & -5.12\mbox{$_{-4}$} & -2.13\mbox{$_{-6}$} &
1.70\mbox{$_{-3}$} & -4.69\mbox{$_{-4}$} & -1.12\mbox{$_{-7}$} &
-4.57135\mbox{$_{-6}$} \\ 
\end{tabular}
\end{ruledtabular}
\end{table*}

In this section we apply the counterpoise correction proposed above to 
calculation of 
interaction-induced first-order properties defined by Eq. (\ref{deltax}). As 
exemplified by the 
recent papers devoted to the helium dimer, this task is considerably more 
challenging than 
computation of the interaction energy alone 
\cite{cencek05,cencek08,patkowski08}. As a benchmark we 
chose as $\hat{X}$  the following operator
\begin{align}
\label{d1}
\hat{D}_1 = \frac{\pi}{2}\alpha^2\sum_a Z_a \sum_i 
\delta(\textbf{r}_{ia}),
\end{align}
where $Z_a$ denotes the nuclear charges, and $\alpha$ is the fine-structure 
constant.  The expectation value of the 
operator $\hat{D}_1$ will be referred to as  the one-electron Darwin correction. 
It appears, 
e.g., in the relativistic Breit-Pauli theory \cite{bethe75} or in calculation   
of hyperfine interactions \cite{Pachucki:11}. Because of the singular character 
of the 
Dirac distribution $\delta(\textbf{r})$ the calculation of the expectation 
value 
of $\hat{D}_1$ is known to be very demanding and slowly convergent with the 
size 
of the Gaussian basis set \cite{piszcz08,przybytek17}.

In Table \ref{d1abs} we show   absolute errors in the interaction-induced 
one-electron Darwin correction for the ground state of H$_2$. The results 
obtained with the asymptotic CP method are not shown in this case because they 
offered no significant improvement over the standard CP.   Our results are 
compared with accurate data of Puchalski et al. 
\cite{puchalski17} whenever available ($R\leq10.0$), and the remaining 
reference 
values are from Ref.~\cite{piszcz08}. The error of the results from 
Refs.~\cite{puchalski17,piszcz08} is negligible in the present context.

The results presented in Table \ref{d1abs} show that the 
counterpoise-corrected method is superior to other techniques in calculation of 
interaction-induced properties.   Similarly as in the previous case, 
cf. Table \ref{intabs}, the counterpoise correction gives  size consistent 
results.  In fact this method is much more effective than in the case of
interaction energy calculations, especially at large $R$.   
Also at small $R$ it performs better than any other 
scheme. For example, at $R=1.40$ (basis 
set $12/300$) the 
errors in the one-electron Darwin correction are about   1.1\% and 0.5\% with 
the pure MC method and 
with subtraction of the exact monomer quantities, respectively. The proposed 
method reduces this 
error to less than 0.08\%. 
It appears that the  CP  method  is particularly well suited for calculation of 
interaction-induced properties with explicitly correlated wave functions.

We close this section by making several observations  concerning the 
computational cost of the proposed 
scheme.  Since in the CP method there is no need to 
reoptimize the nonlinear parameters in the individual basis functions, 
the additional task of constructing  and diagonalizing the  $H_0$  matrix 
adds only a relatively small
contribution to the total cost of ECG calculations 
(dominated by massive nonlinear optimizations).
Once the optimal supermolecular wave function has been obtained, the 
corresponding 
counterpoise-corrected monomer energies become available essentially for free, 
i.e., at a cost of a 
single diagonalization for each $R$. In the asymptotic CP this cost is reduced 
only for a single $R$. We also believe that   existing computer programs for 
explicitly 
correlated calculations can be modified without significant difficulties to incorporate the proposed scheme.

A possible problem related to the calculation of the counterpoise correction is that the basis set 
used for   the diagonalization of $H_0$ is usually a few times larger than that used in diagonalizing 
the dimer  Hamiltonian $H$. 
This might  cause linear dependencies in the basis and,  consequently, problems  in numerical stability of results. We did not observe   this in the calculations presented in this work.
This stable behavior is due to the fact that the basis consists of functions adapted to several 
different representations of the permutation group so the resulting overlap matrix is block diagonal.   
One should note, however, that the $H_0$ matrix is not  only larger, but also  formulas  for its matrix elements are somewhat more complicated than in the case of the  dimer Hamiltonian ($H_0$ does not  commute with all permutations and 
some of them they cannot be moved to only one side of the bracket).  
  
  \section{Conclusions}

In this work we have presented a novel technique, analogous to the counterpoise correction in the 
Boys-Bernardi scheme, to restore size consistency and eliminate basis set superposition error in 
explicitly correlated electronic structure calculations. The new method is based on relaxing the 
Pauli principle in computation of the expectation value of the sum of monomer Hamiltonians (or other 
monomer property operators). This leads to distance dependent monomer energies/properties 
corresponding to the given supermolecular basis set and monomer spin states. It has been shown that 
the proposed method yields interaction energies and interaction-induced properties which vanish 
  at large intermonomer separations.

We would like to stress that the proposed method does \emph{not} provide a way 
to construct individual basis sets for the monomers from a given supermolecular 
basis set. Similarly, the presented method does not allow to calculate 
contributions to Eq. (\ref{deltax}) coming from individual monomers but only 
the sum of monomer quantities.

Exemplary ECG calculations for the hydrogen molecule (H$_2$) indicate that the counterpoise 
correction significantly improves the quality of the results, especially in the 
long-range regions of 
the potential energy curve. This is true  for the interaction energies, but especially for a more 
challenging case of first-order interaction-induced properties (one-electron Darwin correction has 
been tested).  
The additional computational cost of the proposed scheme is small  compared 
to the  necessary optimizations of the supermolecular wave function.

\begin{acknowledgments}
This work was supported by the National Science Center, Poland within the project 
2017/27/B/ST4/02739. The authors are grateful to Jacek Komasa for making his ECG 
program available to us.
\end{acknowledgments}
\vspace{3.0cm}

\bibliography{ref_ecg}

\end{document}